\documentclass[twocolumn,showpacs,prd]{revtex4}

\newcommand{\beq}{\begin{equation}}
\newcommand{\eeq}{\end{equation}}
\newcommand{\bea}{\begin{eqnarray}}
\newcommand{\eea}{\end{eqnarray}}

\usepackage{graphicx}
\usepackage{amssymb}
\usepackage{amsmath}

\usepackage[final]{epsfig}

\begin{document}

\title{A double-domain spectral method for black hole excision data}

\author{Marcus Ansorg}

\affiliation{
 Max-Planck-Institut f\"{u}r Gravitationsphysik \\
(Albert-Einstein-Institut)\\
Am M\"{u}hlenberg 1, D-14476 Golm, Germany}

\date{May 08, 2005}

\begin{abstract}

	In this paper a new double-domain spectral method to compute binary black hole 
	excision initial data is presented. The method solves a system of elliptic partial 
	differential equations in the exterior of two excised spheres. At the surface 
	of these spheres, boundary conditions need to be imposed. 
	As such, the method can be used to construct arbitrary initial data 
	corresponding to binary black holes with specific boundary conditions at 
	their apparent horizons. We give representative examples corresponding to 
	initial data that fulfill the requirements of the quasi-stationary framework, 
	which combines the thin sandwich formulation of the constraint equations with
	the isolated horizon conditions for black holes in quasi-equilibrium. For all 
	examples considered, numerical solutions with extremely high accuracy were obtained 
	with moderate computational effort. Moreover, the method proves to be applicable even 
	when tending toward limiting cases such as large radius ratios for the black holes.

\end{abstract}

\pacs{
04.25.Dm, 
04.20.Ex, 
04.70.Bw  
\quad Preprint number: AEI-2005-102
}

\maketitle

\section{Introduction}\label{Intro}
	
	Spectral methods utilizing multiple spatial domains have been used by 
	many authors in order to solve elliptic partial differential equations in 
	general relativity, 
	see e.g.~\cite{Bonazzola98a,Bonazzola98b,Grandclement02,Gourgoulhon02,Ansorg01b,
	Ansorg:2003br,Ansorg:2002vh,Pfeiffer:2002xz,Pfeiffer:2002wt,Cook:2004}.
	In this paper we demonstrate the strength 
	of these methods by applying them to the calculation of initial data 
	corresponding to binary black hole systems. 
	
	In a `3+1'-splitting of space and time, Einstein's equations may be formulated 
	as constraint and evolution equations. The constraints form elliptic 
	equations on spacelike hypersurfaces, which must be fulfilled by any data 
	set that describes some initial state of the binary system. In constructing 
	data of this kind, two different approaches have been explored, (i) puncture 
	methods \cite{Brandt97b,Beig94,Beig96,Beig:2000ei,Dain00,Ansorg:2004Punc} 
	and (ii) excision techniques \cite{Cook93,Diener99,Hawley:2003az,Thornburg85,Thornburg87,
	Gourgoulhon02,Grandclement02,Pfeiffer:2002wt,Cook:2004, Yo-et-al:2004}.
	
	In the puncture methods, a special 
	pole-like structure of the singularity inside the black hole is assumed which 
	can be taken into account by a specific ansatz for the initial data. Therefore 
	the relevant space for the constraint equations is all of $\mathbb{R}^3$.
	In contrast, the excision techniques 
	solve the constraints only in the exterior of two excised spheroids
	within which the singularities are located. At the surface of these spheroids,
	special boundary conditions need to be imposed, 
	see e.g.~\cite{Misner60,Thornburg85,Thornburg87,Cook:2001wi,Cook:2004,Pfeiffer:2002wt,
	Matzner98a, Grandclement02, Gourgoulhon02}. A promising approach for providing physically 
	realistic data that describe binary systems in quasi-stationary orbits
	is given by the combination of the conformal thin sandwich formulation 
	\cite{york-1999, Pfeiffer-York-2003} (for a review see \cite{cook-2000})
	and the isolated horizon framework 
	\cite{Cook:2001wi,Cook:2004,ashtekar-etal-2000a,dreyer-etal-2003,
	ashtekar-krishnan-2003a,Jaramillo-etal:2004}.
	While the conformal thin sandwich equations incorporate the concept of 
	quasi-stationarity into the constraint equations, the isolated horizon 
	framework describes specific boundary conditions valid at the apparent 
	horizons of the black holes that ensure a quasi-equilibrium state.
	
	In this paper, a new numerical scheme is presented that 
	calculates binary black hole excision data by means of a double-domain 
	spectral method. 
	The scheme is an alternative approach to the spectral techniques used in
	\cite{Grandclement02, Gourgoulhon02} and to those utilized in 
	\cite{Pfeiffer:2002wt,Cook:2004}.
	It works on two spatial subdomains and requires relatively small spectral 
	expansion orders to render highly accurate solutions. The complexity of the 
	numerical scheme is comparable to the one presented in \cite{Ansorg:2004Punc} for a 
	single-domain puncture method, therefore allowing the code to run on a single 
	processor. 

	The central idea of the method is the introduction of two 
	spatial domains, within which the initial data admit a rapidly converging spectral 
	expansion. In order to achieve this it is essential for the data to be smooth 
	within these subdomains. At first, bispherical coordinates \cite{Arfken:1970, Moon:1988} 
	are introduced 
	through which the entire exterior of the two excised balls becomes the image 
	of a single rectangular box (see Section \ref{BisphericalCoordinates}). In 
	particular, a compactification of spatial infinity is realized, which 
	corresponds to a mere line on a side of the box. However, as 
	illustrated in Section \ref{Laplace} for the Laplace equation, a solution to 
	an elliptic equation is in general only ${\cal 	C}^0$ at this line, which 
	suggests the introduction of an additional mapping. This map 
	folds the box along the line in question, see section 
	\ref{Folding}. As a result, for a two-dimensional cross-section, 
	we get a domain of pentangular shape, which we divide 
	up into two quadrangular ones. Each one of the two quadrangular regions is mapped 
	diffeomorphically onto a square. As explained in Section \ref{NumScheme}, 
	spectral expansions for all data entries are carried out within these 
	cuboids, and the collection of constraint equations, boundary, asymptotic 
	fall-off, regularity and  transition conditions (the latter ones to be 
	imposed at the boundary between the two domains) yield a complete set of
	equations to determine all spectral coefficients. We obtain the solution by means of a 
	Newton-Raphson method. For a three-dimensional code, only an iterative scheme 
	for executing the linear step inside the Newton solver is computationally 
	affordable. Together with a reformulation of the regularity conditions (to be 
	enforced at the axis along which the black holes are aligned) a specific 
	plane relaxation scheme has been implemented that results in a convergent 
	iterative procedure. In Section \ref{ExcData} we present first 
	examples satisfying the equations and boundary conditions following from 
	the above quasi-equilibrium framework. We obtain extremely high accuracy 
	even in limiting cases such as very large radius ratios for 
	the excised spheres. Since only relatively little spectral resolution is 
	needed, the code is of low computational cost. This makes it especially 
	useful for the detailed study of wide classes of initial data sets. In 
	Section \ref{Conclusions} we discuss future applications of 
	the scheme, which will include a detailed mathematical and physical 
	investigation of various initial data sets and the use of these data 
	for a dynamical evolution.

\section{Bispherical Coordinates}\label{BisphericalCoordinates}

	\begin{figure*}[t]
		\unitlength1cm
		\epsfig{file=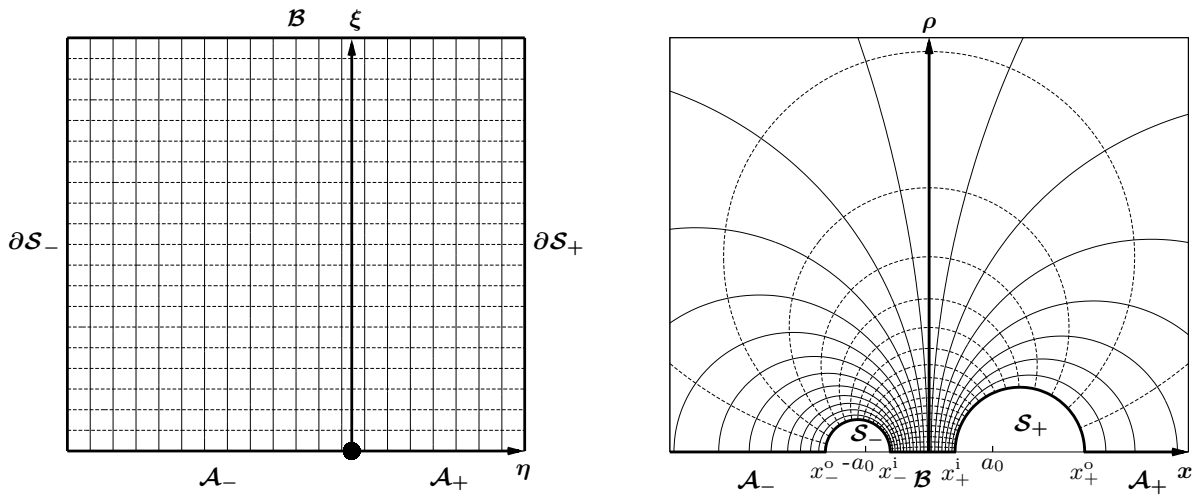,scale=1}
		\caption{
			Illustration of the bispherical mapping 
			(\ref{bispher_complex}-\ref{eta_xi_to_x}) for 
			$D=5\varrho_-, \varrho_+=2\varrho_-$ 
			(with the azimuthal coordinate $\varphi$ suppressed). 
			The entire space exterior to the ${\cal S}_\pm$ in the right panel 
			is obtained as the image of the 
			rectangular box displayed in the left panel. In both panels, solid 
			and dashed curves correspond to constant $\eta$- and 
			$\xi$-coordinate lines, respectively. The line (\ref{infinity}) 
			corresponding to spatial infinity is emphasized by a bullet.
		}
		\label{figure1}
	\end{figure*}

	In this section we consider bispherical coordinates \cite{Arfken:1970, Moon:1988} through which the 
	entire space exterior to the two excised spheres becomes the image of 
	a single rectangular box. Let the spheres be denoted by ${\cal S}_\pm$ and their 
	radii by $\varrho_\pm$. We assume the centers of ${\cal S}_\pm$ to be aligned 
	along the $x$-axis at a distance $D$ from each other 
	and introduce cylindrical coordinates 
	$(x,\rho,\varphi)$ about the $x$-axis, i.~e.
	\beq
		x=x, \quad y=\rho\cos\varphi, \quad z=\rho\sin\varphi \,,
	\eeq
	with $(x,y,z)$ being cartesian coordinates. In the bispherical mapping
	the location of the $(x=0)$-plane is chosen such that the product of the 
	coordinates of inner and outer crossing points of the $x$-axis with the 
	surfaces $\partial{\cal S}_\pm$ (see Fig. \ref{figure1}) is the same 
	for both spheres. That is
	\beq
		x^{\mathrm{i}}_+x^{\mathrm{o}}_+ = x^{\mathrm{i}}_- x^{\mathrm{o}}_- = a_0^2 
	\eeq
	together with
	\bea
		x^{\mathrm{o}}_\pm - x^{\mathrm{i}}_\pm & = & \pm\, 2 \varrho_\pm\,,\quad \\
		(x^{\mathrm{o}}_+ + x^{\mathrm{i}}_+)-(x^{\mathrm{o}}_- + x^{\mathrm{i}}_-) 
												& = & 2 D\,.
	\eea
	Here $a_0>0$ is the common distance of a ``geometric'' center of the two 
	balls to the coordinate origin. It is given by
	\beq
		a_0 =\frac{1}{2}D^{-1}
				\sqrt{D^4-2D^2(\varrho_+^2+\varrho_-^2)
						+(\varrho_+^2-\varrho_-^2)^2}\,.
	\eeq
	The bispherical coordinates $(\eta,\xi,\varphi)$ are most easily introduced by 
	the complex mapping	
	\beq
		\label{bispher_complex}
		c=\mbox{i}a_0\coth\frac{\zeta}{2}\,,
	\eeq	
	where 
	\beq
		c=\rho+\mbox{i}x,\quad \zeta=\eta+\mbox{i}\xi
	\eeq
	are complex combinations
	of cylindrical and bispherical coordinates respectively. Explicitly, the 
	coordinates $\rho$ and $x$ are given in terms of $\eta$ and $\xi$ as follows:
	\bea
		\label{eta_xi_to_rho}
			x &=& a_0\frac{\sinh\eta}{\cosh\eta -\cos\xi} \\
		\label{eta_xi_to_x}
			 \rho  &=& a_0\frac{\sin \xi} {\cosh\eta -\cos\xi}\,.
	\eea
	Note that the angle $\varphi$ remains unchanged under this transformation. 
	
	A characteristic feature of the bispherical coordinates is 
	that all coordinate lines 
	\beq
		\eta=\mathrm{constant}\quad\mbox{and}\quad \xi=\mathrm{constant}
	\eeq
	correspond to circles in the cylindrical coordinates $(x,\rho)$, see Fig. \ref{figure1}.
	In particular it follows that
	\bea
		\label{Circ_xi}
			\left(x-a_0\coth\eta\right)^2+\rho^2 &=&\frac{a_0^2}{\sinh^2\eta}\\
		\label{Circ_eta}
			x^2+\left(\rho-a_0\cot\xi\right)^2   &=&\frac{a_0^2}{\sin^2\xi}\,.
	\eea
	Among these circles we find the surfaces $\partial{\cal S}_\pm$ of the excised 
	spheres, which are described by
	\beq
		\label{eta_pm}
		\eta=\eta_\pm = \pm\mathrm{arsinh}\frac{a_0}{\varrho_\pm}\,.
	\eeq
	From the preceding steps it becomes apparent that the entire space exterior to 
	${\cal S}_\pm$ is obtained as the image of a single rectangular box,
	\beq
		\label{rect_box}
			\eta\in[\eta_-,\eta_+]\,,\quad \xi\in[0,\pi]\,,\quad \varphi\in[0,2\pi)\,.
	\eeq
	While we have already seen that $\eta=\eta_\pm$ corresponds to the surfaces  
	$\partial{\cal S}_\pm$, the faces $\xi=0$ and $\xi=\pi$ 
	(in Fig. \ref{figure1} denoted by ${\cal A}_\pm$ and ${\cal B}$) represent 
	outer and inner sections of the $x$-axis respectively. Note that spatial 
	infinity is obtained as the image of the line
	\beq
		\label{infinity}
			\eta=0=\xi\,,\quad  \varphi\in[0,2\pi)\,.
	\eeq
	At first glance, the bispherical mapping suggests a spectral expansion of the 
	excision data in terms of the coordinates $(\eta,\xi,\varphi)$. However, as 
	will be demonstrated in Section \ref{Laplace}, the data are in general only 
	${\cal C}^0$ at the above line corresponding to spatial infinity. We resolve 
	this issue by another coordinate transformation, as will be discussed in 
	Section \ref{Folding}.

\section{Bispherical solutions of the Laplace-Equation}\label{Laplace}
	
	We include this section about the bispherical solutions of the Laplace equation 
	in order to illuminate some	basic properties of the excision data, in 
	particular 	at spatial infinity. Exterior to ${\cal S}_\pm$ consider
	\beq
		\label{Laplace_U}
		U_{xx}+U_{yy}+U_{zz} =0
	\eeq
	subject to specific Dirichlet conditions that are enforced at the excision 
	boundaries. In bispherical coordinates this reads as follows:
	\bea 
		\label{Laplace_V}
		\lefteqn{a_0^{-2}\left(\cosh\eta -\cos\xi\right)^{5/2}\times }  \\ && \times 
		\nonumber		
		\left(V_{\eta\eta}+V_{\xi\xi}+V_{\varphi\varphi}\csc^2\xi
					+V_\xi\cot\xi-\frac{1}{4}V
		\right)=0\,,
	\eea
	where the potential $V$ is related to $U$ via
	\beq 
		\label{V_To_U}
		U = V\sqrt{\cosh\eta - \cos\xi\;}\,.
	\eeq
	In this formulation the solution can be found by separation of variables \cite{Moon:1988}. 
	It reads
	\bea
		\label{Laplace_Solution}
		\lefteqn{
			U = \sqrt{\cosh \eta - \cos \xi}\; \times
		} \\ && \times \nonumber
		\sum_{l=0}^\infty\sum_{m=-l}^l \left[
			\lambda_{lm} \mathrm{e}^{(l+1/2)\eta}+
			\mu_{lm}     \mathrm{e}^{-(l+1/2)\eta} 
		\right] 
		Y^m_l(\xi,\varphi).
	\eea
	The coefficients $\lambda_{lm}$ and $\mu_{lm}$ are obtained from an expansion 
	of the known Dirichlet values of $V$ at $\eta=\eta_\pm$ with respect to spherical 
	harmonics $Y^m_l$.

	We notice that the analytic behavior of the solution 
	(\ref{Laplace_Solution}) at spatial infinity is determined by the factor
	\bea
		\label{factor}
		\sqrt{\cosh \eta - \cos \xi} &=& 
		\sqrt{2\sinh\frac{\zeta}{2}\sinh\frac{\bar{\zeta}}{2}} \\
		\nonumber          
      	&=& \frac{\sqrt{2}}{2}\sqrt{\eta^2+\xi^2}+{\cal O}(|\zeta|^3)\;
	\eea
	which is only ${\cal C}^0$ at $\zeta=0$. Although one might argue that the 
	decomposition (\ref{V_To_U}) resolves this issue for the Laplace equation, this 
	is not a strategy for solving general nonlinear elliptic equations	since the 
	corresponding solutions contain terms with and without the factor 
	(\ref{factor}).	We therefore pursue a different approach leading to new 
	coordinates in which (\ref{factor}) is analytic.

\section{Folding infinity}\label{Folding}

\begin{figure}[h]
		\unitlength1cm
		\epsfig{file=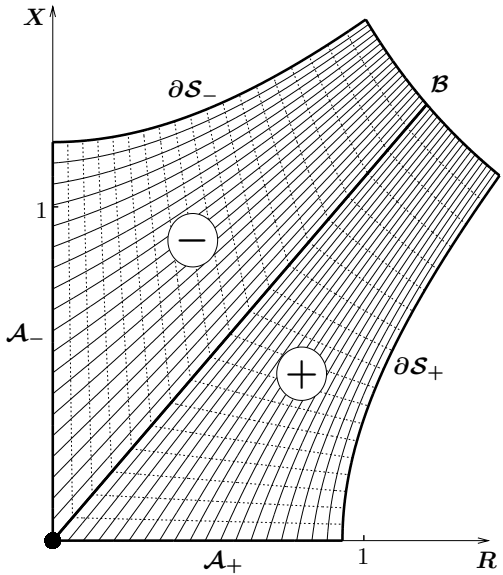,scale=1}
		\caption{
			\label{figure2}
			Illustration of the mapping 
			(\ref{zeta_To_Z}). The example displayed corresponds to the geometrical
			picture shown in Fig. \ref{figure1}. The border that separates the two quadrangular 
			regions $(\pm)$ is a line in this diagram. Solid 
			and dashed curves correspond to constant $A$- and $B$-coordinate lines 
			respectively (see (\ref{AB_To_XR_plus}-\ref{X0R0})).
		}
	\end{figure}

	The regularity problem addressed in the previous section is very closely related 
	to issues discussed in \cite{Ansorg:2004Punc}, see formulae (48-50) therein. Indeed, 
	introducing coordinates $X$ and $R$ via
	\beq
		\label{zeta_To_Z}
		Z=R+\mbox{i}X=\sqrt{\zeta}\;,
	\eeq
	where the square root is taken such that $X$ and $R$ are non-negative in the 
	computational domain (see Fig. \ref{figure2}),	immediately yields that 
	\bea
		\label{factor_XR}
		\lefteqn{
			\sqrt{\cosh \eta - \cos \xi} =\frac{\sqrt{2}}{2} (X^2+R^2)\times
			}
		\\ && \times \nonumber\sqrt{
			\left(1+\sum_{k=1}^\infty\frac{Z^{4k}}{4^k(2k+1)!}\right)
			\left(1+\sum_{k=1}^\infty\frac{\bar{Z}^{4k}}{4^k(2k+1)!}\right)
		}
	\eea
	is analytic in $X$ and $R$, in particular at $X=R=0$.
	
	By performing the transformation (\ref{zeta_To_Z}) and leaving the azimuthal 
	coordinate $\varphi$ unchanged, the rectangular box (\ref{rect_box}) is folded 
	about the line (\ref{infinity}) such that a pentangular region is created upon 
	which $X$ and $R$ are defined, see Fig. \ref{figure2}.
	Since for the spectral method 
	the spectral coordinates $(A,B,\varphi)$ introduced below are given on a 
	cuboid, we draw a border that splits the pentangular region into 
	two	subdomains $(\pm$) of quadrangular shape. Now the coordinates 
	$(A,B)$	with
	\beq
		(A,B)\in[0,1]\times[0,1]
	\eeq
	are mapped to values $(X,R)$ on these subdomains via \\ \\
	$\bullet$ region $+$:
	\beq
		\label{AB_To_XR_plus}
		\begin{array}{ccc}
			X &=& -X^+ AB +R_+ A \sinh(\mu^+ B) + X_0 B \mathrm{e}^{\nu^+ A} 
			\\[2mm]
			R &=& -R^+ AB +R_+ A \cosh(\mu^+ B) + R_0 B \mathrm{e}^{-\nu^+ A} 
		\end{array}
	\eeq
	$\bullet$ region $-$:
	\beq
		\label{AB_To_XR_minus}
		\begin{array}{ccc}
			X &=& -X^- AB +X_- A \cosh(\mu^- B) + X_0 B \mathrm{e}^{\nu^- A} 
			\\[2mm]
			R &=& -R^- AB +X_- A \sinh(\mu^- B) + R_0 B \mathrm{e}^{-\nu^- A} 
		\end{array}
	\eeq
	where
	\beq	
		\label{parameters}
		\begin{array} {lcl} 
			Z_\pm   &=& R_\pm+\mbox{i}X_\pm = \sqrt{\eta_\pm} \;,\\[2mm]
			Z^\pm   &=& R^\pm+\mbox{i}X^\pm = \sqrt{\eta_\pm+\mbox{i}\pi}\;,\\[5mm]
			\mu^+   &=& \mathrm{arsinh}(X^+/R_-)\;,\\[2mm]
			\mu^-   &=& \mathrm{arsinh}(R^-/X_-)\;,\\[2mm]
			\nu^\pm &=& \ln(X^\pm/X_0) \;.
		\end{array}
	\eeq
	A value for $X_0$ ($R_0=\pi/(2X_0)$) may be chosen in order 
	to create subdomains of comparable size:
	\beq
		\label{X0R0}
		X_0=\sqrt{X^-X^+}\;, \quad R_0=\sqrt{R^-R^+}\;.
	\eeq
	With this transformation, the surfaces $\partial{\cal S}_\pm$ are obtained 
	for $A=1$. Exterior and interior sections of the $x$-axis 
	(i.e.~${\cal A}_\pm$ and ${\cal B}$) are given by $B=0$ 
	and $B=1$ respectively, and the common border between the two 
	subdomains is described by $A=0$. 
	
\section{The numerical scheme}\label{NumScheme}

	Our numerical scheme possesses a number of common features with the method 
	described in \cite{Ansorg:2004Punc} for the spectral calculation of 
	puncture initial data.
	However, there are also specific differences,
	which we depict in the following paragraphs.

	In a spectral approximation all functions $U^{\kappa}$
	to be determined by our elliptic boundary value problem 
	are considered at specific gridpoints that correspond to the 
	zeros or extrema of the spectral basis functions being used. 
	Here we perform Chebyshev expansions with respect to $A$ 
	and $B$ and a Fourier expansion with respect to $\varphi$. 
	Since the quasi-stationary framework requires sophisticated boundary values 
	at $\partial{\cal S}_\pm$ for the initial data (see Section \ref{ExcData}), 
	we choose to use the extrema of the Chebyshev polynomials
	so as to have gridpoints lying on the boundary. 
	That is, the gridpoints are given by 
	\bea
		\nonumber 	A_j&=&\sin^2\left[\frac{\pi j}{2(n_A-1)}\right] , \\
		\label{Gridpoints}
					B_k&=&\sin^2\left[\frac{\pi k}{2(n_B-1)}\right] , \\[2mm]
		\nonumber 	\varphi_l&=&\frac{2\pi l}{n_\varphi}
	\eea
	where 
	\beq
		\label{jkl} 0\leq j<n_A,\quad 0\leq k<n_B,\quad 0\leq l<n_\varphi\,.
	\eeq 
	The numbers $n_A, n_B$ and $n_\varphi$ describe the spectral expansion orders 
	of our scheme. 

	The next step in the scheme is the setting of a vector $\vec{U}$, 
	whose components are derived from the values
	\beq
		\label{U_values}
		U^{\kappa\,\pm}_{jkl}
	\eeq
	of all functions $U^{\kappa}$ at the above gridpoints $(A_j,B_k,\varphi_l)$ in 
	the two subdomains $(\pm)$. The index $\kappa\in\{1,\ldots,\kappa_0\}$ 
	labels the different functions appearing in the elliptic system, 
	with $\kappa_0$ being the total number of equations ($\kappa_0=5$ for the 
	quasi-stationary constraints discussed in Section \ref{ExcData}). 
	
	A way to set $\vec{U}$ that works well in the context of the
	relaxation method described below is given by the following unique decomposition 
	of $U^{\kappa}$\,:
	\bea
		\label{decomp_U}
		U^\kappa(A,B,\varphi)&=&V^\kappa(A,B)+W^\kappa(A,B,\varphi)\,,\\
		\label{W_eq_0}
		W^\kappa(A,B,0)      &=&0\,.
	\eea 
	We use this decomposition and collect all values $V^{\kappa\,\pm}_{jk}$ 
	and $W^{\kappa\,\pm}_{jkl}$ for $l>0$ in order to build 
	up the vector $\vec{U}$. This means that in our approach the 
	$U^{\kappa\,\pm}_{jkl}$ are not stored directly in $\vec{U}$, but can be recovered
	from the entries of $\vec{U}$ through the above sum (\ref{decomp_U}).

	The collection of elliptic equations valid in the exterior of ${\cal S}_\pm$, 
	boundary conditions imposed at $A=1$ (i.e.~at $\partial{\cal S}_\pm$), 
	transition conditions to be imposed at $A=0$
	(i.e.~at the common border) and specific regularity conditions 
	that need to be fulfilled at $B=0$ and $B=1$ (i.e.~at the $x$-axis) yield a 
	discrete non-linear system
	\beq
		\label{f_of_U}
		\vec{f}(\vec{U})=0
	\eeq
	of dimension 
	\beq
		\label{total_dim}
		2\kappa_0 n_A n_B n_\varphi\,.
	\eeq 
	The corresponding solution describes the desired spectral approximation of the
	solution to our elliptic boundary value problem. 
	Let us discuss in detail the several entries 
	$f^{\kappa\,\pm}_{jkl}$ of the vectorial function $\vec{f}$:
	\begin{enumerate}
		\item 
			The spectral method enables us to calculate first and second 
			derivatives of the $U^\kappa$ from the values 
			$U^{\kappa\,\pm}_{ijk}$ at the above grid points
			up to the chosen approximation order. Using them in the given elliptic
			equations evaluated at these grid points provides relations 
			between the $U^{\kappa\,\pm}_{jkl}$. Note that the elliptic equations are 
			only considered at inner grid points, i.e.~for 
			\bea
				\nonumber   				1&\leq j\leq& n_A-2 \,,\\
				\label{which_inner_points}	1&\leq k\leq& n_B-2 \,,\\
				\nonumber					0&\leq l<&n_\varphi\,, 
			\eea
			thus filling 
			\beq
				\label{number_inner_points}
					2\kappa_0 (n_A-2) (n_B-2) n_\varphi
			\eeq
			entries of $\vec{f}$.
		\item
			In exactly the same manner, we use the boundary conditions valid at 
			$A=1$ to fill
			\beq
				\label{number_boundary_points}
					2\kappa_0 n_B n_\varphi
			\eeq
			entries of $\vec{f}$. Note that the boundary conditions are also considered
			at the $x$-axis, i.e.~at gridpoints with $j=0$ or $j=n_B-1$.
		\item
			Since the border between the two domains is some artificial transition surface 
			that we have introduced, each function $U^\kappa$ must 
			be smooth there. In the formulation of the elliptic boundary value 
			problem, this is ensured by requiring the $U^\kappa$ to be continuous 
			and to possess continuous normal derivatives at $A=0$. These conditions
			are enforced along the entire border except the line 
			(\ref{infinity}) where the values of $U^\kappa$ at infinity are 
			imposed, thus leading altogether to
			\beq
				\label{number_transition_points}
					2\kappa_0 n_B n_\varphi
			\eeq
			entries of $\vec{f}$. 
		\item
			Special care is needed for setting the remaining discrete equations 
			that correspond to regularity conditions along the $x$-axis. Any function
			$U$ that is smooth along the $x$-axis with respect to cartesian 
			coordinates can be expanded in a Taylor series:
			\bea
				\label{Taylor}
				\lefteqn{\quad U(x,y,z)=U(x,0,0)\,+}\\[2mm]
				&&\nonumber +\rho[U_y(x,0,0)\cos\varphi
						+\,U_z(x,0,0)\sin\varphi]+{\cal O}(\rho^2)
			\eea
			from which it follows that
			\beq
				\label{Axis_1}
				\lim_{\rho\to 0}U_{\varphi\varphi} = 0 
			\eeq
			and
			\beq
				\label{Axis_2}
				\lim_{\rho\to 0}(U_{\rho}+U_{\rho\varphi\varphi}) = 0\;. 
			\eeq
			The entire set of equations turns out to be solvable only if we consider 
			both conditions (\ref{Axis_1}) and (\ref{Axis_2}). In particular, we 
			require (\ref{Axis_1}) at gridpoints with $\varphi>0$ (i.e.~$l>0$) 
			and (\ref{Axis_2}) at gridpoints with  $\varphi=0$ (i.e.~$l=0$). 
			The corresponding relations fill the remaining
			\beq
				\label{number_axis_points}
					4\kappa_0 (n_A-2) n_\varphi
			\eeq
			entries of $\vec{f}$. Note that different conditions need to be imposed
			if the underlying function $U$ is not smooth along the $x$-axis with 
			respect to cartesian coordinates. An example for this is given by the 
			bispherical components of a vectorial function. 
			In a subsequent publication we will
			discuss in detail this issue in connection with the momentum constraint 
			equations.
	\end{enumerate}
	In treating the system (\ref{f_of_U}) we follow very much the approach described 
	in \cite{Ansorg:2004Punc}, see section II therein. In particular, the solution is obtained by 
	Newton-Raphson iterations, and the linear step inside this solver is performed 
	with the preconditioned `Biconjugate Gradient Stabilized (BCGSTAB)' method
	\cite{Barrett93}. In complete analogy to \cite{Ansorg:2004Punc}, we construct 
	a second order finite difference representation of the Jacobian of 
	(\ref{f_of_U}), which will be used in the preconditioning step. Note that, because 
	of the decomposition (\ref{decomp_U}), for every gridpoint not only adjacent 
	neighboring points need to be considered, 
	but also the values for $l=0$. The resulting finite difference Jacobian matrix 
	has therefore at most $27\kappa_0$ non-vanishing entries per row.

	A crucial difference to the method used in \cite{Ansorg:2004Punc} is the performance of 
	the preconditioning step in which an approximate solution of the system
	\beq
		\label{FD_Jacobian}
		J_{\mathrm{FD}}\,\delta\vec{U} = -\vec{f}(\vec{U}) 
	\eeq
	is obtained. The matrix $J_{\mathrm{FD}}$ appearing here
	represents the finite difference approximation of the Jacobian
	$J=\partial\vec{f}/\partial\vec{U}$ which was mentioned above. 
	The vector $\delta\vec{U}$
	describes a correction to $\vec{U}$ obtained through the values
	$\vec{f}(\vec{U})$. 

	We use a specific preconditioner that consists 
	of successive plane relaxations with respect to the planes 
	$\varphi=\mbox{constant}=\varphi_l$. 
	These plane relaxations in turn are composed of two different 
	line relaxation schemes. 
	We consider the method in more detail:
	\begin{enumerate}
		\item 
			\label{linerelax1}
			For the first one of the two line relaxation schemes, 
			let us pick one of the two regions ($\pm$) and 
			some values $j,l$ and $\kappa$. 
			Then all values $\delta W^{\kappa\,\pm}_{jkl}$ 
			(or $\delta V^{\kappa\,\pm}_{jk}$ 
			for $l=0$) along the coordinate line 
			\beq
				\label{line_relax_B}
				A=A_j,\, B\in[0,1],\, \varphi=\varphi_l
			\eeq
			are determined simultaneously through (\ref{FD_Jacobian})
			while all other entries of 
			$\delta\vec{U}$ are held fixed. 
			This gives a tridiagonal system of dimension $n_B$ 
			which can be solved with low computational cost. 

			We perform this procedure within the plane $\varphi=\varphi_l$
			for all values $j$ and $\kappa$ and in both regions ($\pm$).
			Hereby, the loop over $j$ takes on all even 
			values first and then all odd ones, thus describing a `zebra' 
			line relaxation scheme.
		\item 
			\label{linerelax2}
			Similarly, for the second line relaxation scheme $k,l$ and $\kappa$ 
			are chosen, and all values $\delta W^{\kappa\,\pm}_{jkl}$ 
			(or $\delta V^{\kappa\,\pm}_{jk}$ for $l=0$) along the coordinate line 
			\beq
				\label{line_relax_A}				
				A\in[0,1),\, B=B_k,\, \varphi=\varphi_l
			\eeq
			are determined simultaneously
			through (\ref{FD_Jacobian}) in both regions ($\pm$), thus leading 
			to a tridiagonal system of dimension ($2n_A-3$).
			Here we make use of the continuity
			of the $U^\kappa$ at the transition border. Note that we leave out
			the values at the boundary $A=1$, which for specific boundary conditions 
			might be problematic for the convergence of the relaxation.
			We therefore obtain an update of these values only through the 
			line relaxations described in \ref{linerelax1}. 
			
			Again we apply these steps within the plane $\varphi=\varphi_l$ 
			for all values $k$ and $\kappa$ using a zebra loop with respect to $k$. 
		\item 
			The plane relaxation scheme within the plane 
			$\varphi=\varphi_l$ is composed 
			of several of the above orthogonal line relaxation schemes. A zebra loop 
			with respect to $l$ completes the relaxation method.
	\end{enumerate}
	The preconditioning step, consisting of a number of such relaxations
	(a typical value is 20), is the fragile ingredient of the method. A simple 
	reformulation of the boundary or regularity conditions might spoil the 
	convergence of the scheme that otherwise is obtained through its iterative
	application. Since for interesting excision data the collection of elliptic 
	equations and boundary conditions form a complicated system, it is very 
	difficult to know how to pick a particular formulation in advance 
	in order to ensure a stable convergent iteration. Nevertheless, for the 
	relevant examples discussed in Section \ref{ExcData} it was possible to cast 
	the conditions into a suitable form.
	
	Note that in general, relaxation schemes are computationally quite expensive 
	if high resolution is used. In this case, it would be possible to 
	include the relaxation scheme within a multigrid solver. However, since the 
	spectral resolution can be chosen to be rather small (see Section \ref{ExcData}), 
	the method is already very efficient as it stands.

	A preconditiong step consisting of 20 relaxation iterations gives
	a good approximation for the `BCGSTAB'-method. Typically, only about 6 
	iterations within this scheme are needed to complete the linear Newton step. 
	Thus the numerical scheme converges rapidly to the desired finite spectral
	approximation of the solution. 

\section{Quasi-stationary excision data}\label{ExcData}

	The numerical scheme presented is applicable to an arbitrary
	set of elliptic equations that is valid in the exterior of two spheres
	and subject to specific boundary conditions required at the surfaces of these shells.
	In particular, it should prove fruitful for the calculation of a variety of
	initial data sets corresponding to binary black hole systems. In this section
	we apply the method to the important example, in which the initial data are 
	given through the quasi-stationary framework. 

	In a first subsection we review the thin sandwich approach which yields the 
	set of elliptic equations to be considered. Next the isolated horizon 
	boundary conditions describing a black hole in a quasi-equilibrium 
	state are discussed.
	The main focus of this section is on the presentation of exemplary solutions 
	to the corresponding boundary value problem, which have been obtained by means of
	the scheme to an extremely high accuracy. In particular, we illustrate the strength 
	of the method in the limiting case of very large ratios $\varrho_+/\varrho_-$. 

	\subsection{The thin sandwich equations}

		In the ADM-formulation of a `3+1'-splitting of the spacetime
		manifold, the general relativistic line element is written as
		\beq
			\label{line_element}
			ds^2=\gamma_{ij}(dx^i+\beta^i dt)(dx^j+\beta^j dt) -\alpha^2 dt^2\,,
		\eeq
		where $\gamma_{ij}$ is the 3-metric, and $\beta^i$ and $\alpha$ are the shift
		vector and lapse function respectively. Einstein's field equations
		can be split up into a set of constraint and evolution equations for the 12 
		quantities $\gamma_{ij}$ and $K_{ij}$ where
		\beq
			\label{extr_curv}
			K_{ij}=\frac{1}{2\alpha}\left(
				\bar{\nabla}_i\beta_j+\bar{\nabla}_j\beta_i
					-\partial_t\gamma_{ij}\right)
		\eeq
		is the extrinsic curvature ($\bar{\nabla}_j$ represents the spatial covariant 
		derivative associated with $\gamma_{ij}$). The definition
		(\ref{extr_curv}) yields one of two sets of six evolution equations. 
 
		
		In York's `Conformal Thin-Sandwich Decomposition' \cite{york-1999} the evolution of the metric 
		between two neighboring slices $t=\mathrm{constant}$ is considered. More precisely, 
		the evolution equations (\ref{extr_curv}) make it possible to 
		prescribe specific values 
		for the initial time derivative of the conformal metric $\tilde{\gamma}_{ij}$, 
		which is connected to $\gamma_{ij}$ via 
		$\gamma_{ij}=\psi^4\tilde{\gamma}_{ij}$ where $\psi$ is a 
		`conformal factor'. Note that this splitting is unique only if we 
		require some normalization for 
		$\tilde{\gamma}_{ij}$, e.g.~ $\mathrm{det}(\tilde{\gamma}_{ij})=1$. 
		In addition the extrinsic curvature is decomposed:
		\beq
			\label{decomp_Kij}
			K^{ij}=\psi^{-10}\tilde{A}^{ij}+\frac{1}{3}\gamma^{ij}K\,,
				\quad K=K^i_i\,.
		\eeq

		In order to obtain initial data corresponding to a binary black hole in a 
		quasi-stationary orbit, the constraint and evolution equations are considered
		in a comoving frame of reference in which the time derivatives of the metric
		quantities are assumed to be small initially. The thin sandwich decomposition 
		allows us to set 
		\beq
			\partial_t\tilde{\gamma}_{ij}=0
		\eeq 
		in an initial slice $t=t_0$ from which by virtue of (\ref{extr_curv}) it follows that
		\beq
			\label{A_of_beta}
			\tilde{A}_{ij}=\frac{\psi^6}{2\alpha}(\tilde{\mathbb{L}}\beta)_{ij}
		\eeq
		with
		\beq
			(\tilde{\mathbb{L}}\beta)_{ij} = 
				\tilde{\nabla}_i\beta_j+\tilde{\nabla}_j\beta_i 
					-\frac{2}{3}\tilde{\gamma}_{ij}\tilde{\nabla}_k\beta^k\,.
		\eeq
		In this formulation the constraint equations are given by
		\bea
			\hspace{-1cm}\label{Hamiltonian}
			\tilde{\nabla}^2\psi-\frac{1}{8}\psi\tilde{R}-\frac{1}{12}\psi^5K^2
				+\frac{1}{8}\psi^{-7}\tilde{A}_{ij}\tilde{A}^{ij}&=&0 \\
			\hspace{-1cm}\label{Momentum}
			\tilde{\nabla}_j(\tilde{\mathbb{L}}\beta)^{ij}
				-(\tilde{\mathbb{L}}\beta)^{ij}\tilde{\nabla}_j(\ln\alpha\psi^{-6})
				-\frac{4}{3}\alpha\tilde{\nabla}^i K&=& 0\,.
		\eea
		Note that in these formulae $\tilde{\nabla}_j$ and $\tilde{R}$ are the spatial
		covariant derivative and the Ricci scalar respectively, associated with the
		conformal metric $\tilde{\gamma}_{ij}$. Indices are raised and lowered with 
		respect to this metric. 

		The Hamiltonian (\ref{Hamiltonian}) and momentum constraints (\ref{Momentum}) 
		are elliptic equations which can be used to determine the conformal factor
		$\psi$ and the shift $\beta^i$ respectively. Moreover, in the quasi-stationary
		framework it is possible to consider an additional equation which
		is obtained from the evolution equations through the requirement
		\beq
			\partial_t K=0
		\eeq 
		on the initial slice. This also gives an elliptic equation which determines
		the lapse function $\alpha$:
		\bea
			\hspace{-1cm}\label{dot_K}
			\lefteqn{
				\tilde{\nabla}^2(\alpha\psi) = } \\
			&& \nonumber \alpha\psi\left[\frac{1}{8}\tilde{R}+\frac{5}{12}\psi^5K^2
				+\frac{7}{8}\psi^{-8}\tilde{A}_{ij}\tilde{A}^{ij} \right] 
				+\psi^5\beta^i\tilde{\nabla}_i K \,.
		\eea
		Thus in total we have five elliptic equations (\ref{Hamiltonian}, 
		\ref{Momentum}, \ref{dot_K}) in order to determine the five quantities
		$\psi,\beta^i$ and $\alpha$. Note that the conformal metric
		$\tilde{\gamma}_{ij}$ and the trace $K$ are free data that enter these equations.
		From the framework being presented no further restrictions on these
		quantities follow. In this paper we choose examples that correspond to maximal and
		conformally flat data, i.e.~we set 
		\beq
			\label{Maximal} K=0
		\eeq
		 and 
		\beq
			\label{ConFlat} \tilde{\gamma}_{ij}=\delta_{ij}
		\eeq
		in the cartesian coordinates $(x,y,z)$ (with $\delta_{ij}$ denoting the Kronecker symbol).

	\subsection{Boundary conditions}

		The above quasi-stationary formulation is completed by a set of boundary 
		conditions that control the data at the excision boundaries and at infinity.
		The excision boundary conditions are given through the 
		`Isolated Horizon Framework' \cite{Cook:2001wi,Cook:2004,ashtekar-etal-2000a,
		dreyer-etal-2003, ashtekar-krishnan-2003a,Jaramillo-etal:2004} (see also
		\cite{Grandclement02,Gourgoulhon02})
		and describe black holes in a quasi-equilibrium state. 
		In particular the following is required:
		\begin{enumerate}
			\item 
				Within the initial slice, both excision boundaries are apparent horizons, 
				i.e.~two-dimensional hypersurfaces with $S^2$ topology and the property that 
				the outgoing null vectors $k$ possess vanishing expansion.
			\item
				Initially, the apparent horizon is tracked along $k$ and
				its coordinate location does not move in the time evolution of 
				the data.
		\end{enumerate}
		Cook \cite{Cook:2004, Cook:2001wi} incorporated these requirements into 
		the thin sandwich formulation and arrived at the following boundary conditions 
		for $\psi$ and $\beta^i$ that are required at $\partial{\cal S}_\pm$:
		\bea
			\label{psi_bound}
			\tilde{s}^i\tilde{\nabla}_i(\ln\psi) &=& 
				-\frac{1}{4}(\tilde{h}^{ij}\tilde{\nabla}_i\tilde{s}_j
				- \psi^2 J) \\ 
			\label{beta_bound}
			\beta^i&=&\alpha\psi^{-2}\tilde{s}^i+\beta_{||}^i\,.
		\eea
		Here the vector $\tilde{s}^i$ is the outward pointing unit vector
		normal to $\partial{\cal S}_\pm$ (with respect to $\tilde{\gamma}_{ij}$)
		and $\tilde{h}_{ij}=\tilde{\gamma}_{ij}-\tilde{s}_i\tilde{s}_j$ the
		conformal metric induced on $\partial{\cal S}_\pm$. The quantity $J$
		is given by 
		\beq
			\label{J}
			J=\psi^6\tilde{h}_{ij}\tilde{A}^{ij}+\frac{2}{3}K
		\eeq
		and the vector $\beta_{||}^i$ tangent to $\partial{\cal S}_\pm$ is 
		proportional to a conformal Killing vector
		of the conformal metric $\tilde{h}^{ij}$, with the proportional factor
		describing the angular velocity of the black hole. Note that the 
		isolated horizon framework does not yield a boundary condition for the lapse
		$\alpha$, which we are therefore free to choose. 

		In the examples discussed below we consider corotational black holes 
		for which the angular velocity of the black holes, as seen within the 
		comoving frame of reference, vanishes, i.e.
		\beq
			\label{CoRot}\beta_{||}^i=0\,.
		\eeq
		Moreover, we take the following boundary condition for the lapse, which was used
		among others in \cite{Cook:2004}\,:
		\beq
			\label{alpha_bound}
			\tilde{s}^i\tilde{\nabla}_i(\alpha\psi) =0 \,.
		\eeq
		
		At infinity we impose the appropriate asymptotic behavior 
		\bea
			\label{alpsi_at_inf}
				\lim_{|\vec{r}|\to\infty}\psi=\lim_{|\vec{r}|\to\infty}\alpha&=&1 \\
			\label{beta_at_inf}
				\lim_{|\vec{r}|\to\infty}[\beta^i-(\vec{\Omega}\times\vec{r}\,)^i]&=&0
		\eea
		resulting from the fact that $\beta^i$ is a shift vector in a comoving 
		frame of reference that rotates with the angular velocity $\vec{\Omega}$.
		In the cartesian coordinates $\vec{r}=(x,y,z)$ we take 
		\beq
			\label{Omega}\vec{\Omega} = \Omega\vec{e}_z
		\eeq
		corresponding to an orbital motion of the black holes.

	\subsection{Examples}\label{Examples}

	A well known maximal and conformally flat solution to the above 
	quasi-stationary boundary value problem is given by the Misner-Lindquist initial data 
	\cite{Misner60, lindquist63} which are used for the evolution of a head-on collision of 
	two black holes. These data are characterized by vanishing $\alpha$ and $\beta_{||}^i$ 
	at the horizon (i.e.~these data do not fulfill the condition (\ref{alpha_bound}) which we 
	impose below) and moreover by $\Omega=0$ (from which it naturally follows that they 
	do not represent two orbiting black holes in a quasi-stationary state). 
	In order to find quasi-stationary data, we need additional requirements 
	such as the equality of ADM and Komar masses discussed below. Nevertheless, we start 
	the investigation of the quasi-stationary boundary value problem in question by 
	calculating solutions that represent 
	modifications to the Misner-Lindquist data. In particular, we take the boundary condition 
	(\ref{alpha_bound}) (instead of $\alpha=0$) but retain all remaining conditions. 
	Whereas the corresponding boundary value problem 
	for a single black hole admits a solution which can be given
	explicitly \cite{estabrook-etal-73, Cook:2004}, 
	no exact solution is known in the binary case. 
	However, we may take a superposition of the solutions known for the single black 
	hole in order to create `start up' data for the Newton-Raphson scheme inside 
	the numerical scheme. We consider the resulting solutions for $D=10\varrho_-$ and 
	four different choices of the radius ratio, $\varrho_+/\varrho_-\in\{1;10;100;1000\}$. 
	The excellent convergence of the spectral method has been checked globally 
	for all functions $U^\kappa$ appearing in our elliptic boundary value problem.
	We illustrate it in Fig. \ref{figure3} for the total ADM-mass of 
	the system which is given in Cartesian coordinates by the following surface 
	integral evaluated at infinity:
	\beq
		\label{ADM_Mass}
			M_{\mathrm{ADM}}=\frac{1}{16\pi}\int\limits_\infty 
				(\gamma_{ij,j}-\gamma_{jj,i}) d^{\,2}S_i\,.
	\eeq
	We find that the convergence seems to be geometric, as exhibited by a roughly
	linear decrease of the error in this diagram.

	Note that almost machine accuracy is reached for all ratios considered, thus 
	proving that the method is well suited to the case of extreme radius ratios 
	(and likewise to situations in which the distance $D$ is extremely large). 
	This may be clarified by the following considerations: 

	\begin{figure}[h]
		\epsfig{file=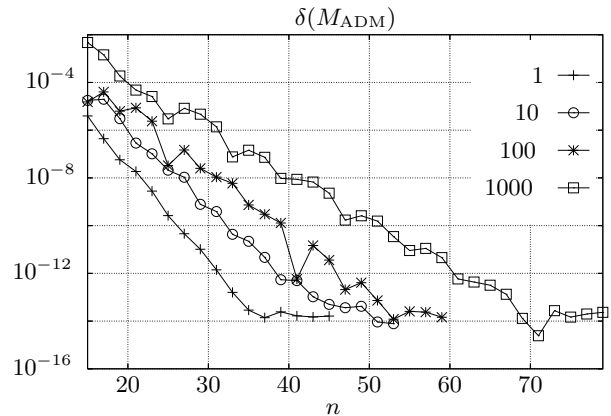,scale=1}
		\caption{\label{figure3}
			The convergence of the ADM-mass corresponding to 
			binary black hole initial data sets
			with vanishing orbital angular velocity $\Omega=0$. The geometrical
			parameters of the individual configurations are given by $D=10\varrho_-$ and 
			$\varrho_+/\varrho_-\in\{1;10;100;1000\}$. For these axisymmetric
			calculations $n_A=n_B=n$ and $n_\varphi=3$ have been chosen. 
			We compared the corresponding results for the ADM-masses to those 
			of reference solutions with $n\in\{50;60;70;80\}$ for the above choices of 
			$\varrho_+/\varrho_-$.
			}
	\end{figure}
	
	We know from the study of single black holes that 
	the Dirichlet boundary values of the data remain restricted independent 
	of the ratio $\varrho_+/\varrho_-$. However, because of the different scalings
	involved in the problem, one might not expect that the normal derivatives 
	of the data at the surface of the small sphere remain bounded as 
	$\varrho_+/\varrho_-$ increases. This in turn
	would mean that a spectral expansion needs more and more resolution, which
	for extreme ratios becomes computationally unachievable. 

	To illustrate why nevertheless our method handles these critical cases with
	moderate computational effort, consider a special
	solution (\ref{Laplace_Solution}) to the Laplace equation (\ref{Laplace_V}), 
	given by
	\beq
		V=\mathrm{e}^{(\eta_- - \eta)/2}\,.
	\eeq
	It corresponds to the potential of a single point mass located at 
	$(x,\rho)=(-a_0,0)$ and possesses a finite value $V=1$ at ${\cal S}_-$. 
	For the potential $V$ we need not fold the rectangular box (\ref{rect_box})
	at the line (\ref{infinity}) in order to obtain analyticity there, 
	but may directly introduce spectral coordinates on (\ref{rect_box}), 
	i.e.~$(\hat{A},\hat{B})$ via 
	\beq
		\eta=\eta_- + (\eta_+-\eta_-)\hat{A}\,,\qquad\xi=\pi\hat{B}\,.
	\eeq
	Now the derivative 
	\beq
		\left.\frac{\partial V}{\partial\hat{A}}\right|_{\hat{A}=0}=-
			\frac{1}{2}(\eta_+-\eta_-)
	\eeq
	tends only logarithmically to infinity as $\varrho_-/a_0\to 0$ 
	(see (\ref{eta_pm})). 
	Since the folding of (\ref{rect_box}) along (\ref{infinity}) does not 
	modify the qualitative behavior of the derivatives, 
	a rapid spectral convergence emerges generally in these critical cases.
	Note that our coordinates are somewhat similar to those introduced in 
	\cite{Ansorg-Petroff-2005} via
	a logarithmic mapping in order to capture very different length scales 
	appearing there.

	In the second example we calculate initial data corresponding to two 
	corotational black holes in a quasi-stationary orbit. It has been argued 
	\cite{Grandclement02, Gourgoulhon02, Cook:2004, Cook:2001wi} that 
	a suitable value for the angular velocity $\Omega$ is obtained by requiring
	the equality of the ADM-mass and the Komar mass which is defined by the 
	following surface integral at infinity:
	\beq
		\label{Komar_Mass}
			M_{\mathrm{K}}=\frac{1}{4\pi}\int\limits_\infty 
				\gamma^{ij}(\bar{\nabla}_i\alpha-\beta^k K_{ik}) d^{\,2}S_j \,.
	\eeq
	In the example considered here, $D=10\varrho_-$ and $\varrho_+=\varrho_-$
	have been chosen. For these data an orbital angular velocity 
	$\Omega\approx0.036975/\varrho_-$ emerges. Note that we introduce the vector 
	\beq
		\label{hat_beta}
		\hat{\beta}^i=\beta^i-(\vec{\Omega}\times\vec{r})^i
	\eeq
	into our
	numerical scheme which enables us to set definite 
	values of the corresponding $U^{\kappa}$ at infinity.
	
	\begin{figure}[h]
		\epsfig{file=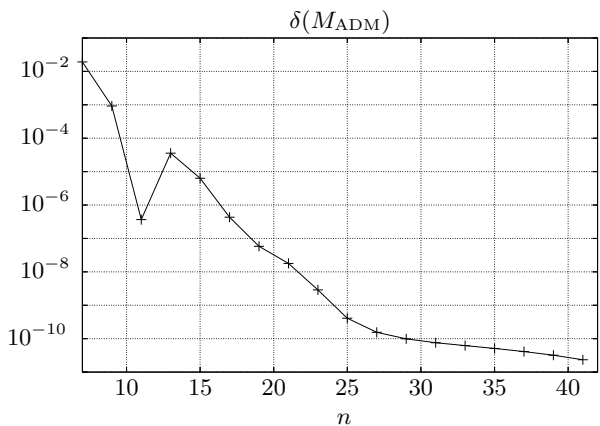,scale=1}
		\caption{\label{figure4}
			The convergence of the ADM-mass corresponding to a 
			corotating binary black hole initial data set in a 
			quasi-stationary orbit. The geometrical
			parameters of the configuration are given by $D=10\varrho_-$ and 
			$\varrho_+=\varrho_-$. The data are characterized by the equality 
			$M_{\mathrm{ADM}}=M_{\mathrm{K}}$ through which an 
			orbital angular velocity of $\Omega\approx0.036975/\varrho_-$
			emerges. For these calculations the 
			spectral resolutions $n_A=n_B=2n_\varphi+1=n$ have been chosen.
			We compared the corresponding results for the ADM-mass to that 
			of the reference solution with $n=51$.
		}
	\end{figure}
 
	Again we obtain a rapid overall convergence of the method which we
	illustrate by displaying the relative error of the total ADM-mass, 
	see Fig. (\ref{figure4}). In contrast to the above axisymmetric examples, 
	a roughly linear section of the curve for small resolution is followed 
	by a much more flattened part as the accuracy approaches 10 digits.
	A similar convergence rate was found for puncture initial data possessing 
	individual linear momenta of the black 
	holes \cite{Ansorg:2004Punc}. We plan to discuss mathematical issues of this 
	kind thoroughly in a subsequent publication.
	\\[1cm]

\section{Conclusions}\label{Conclusions}

	In this paper we presented a numerical scheme that calculates
	binary black hole excision data by means of spectral methods. The central idea
	of the scheme is the introduction of specific coordinates that are related
	to bispherical coordinates, in order to permit rapid convergence of the
	spectral expansions. In particular, the entire space exterior to the two black
	holes is obtained as the image of two spatial domains within which the spectral
	expansions are carried out. A special formulation of the boundary and regularity
	conditions enables us to use a specific iterative technique to approach the solution
	that corresponds to the spectral approximation of the desired data. 

	The scheme has been used to calculate examples corresponding to the 
	important quasi-stationary framework which is given by the conformal thin-sandwich 
	decomposition 
	together with the isolated horizon boundary conditions. It exhibits a rapid spectral 
	convergence of the numerical solutions up to an extremely high accuracy. 
	In particular, configurations with large radius ratios 
	of the black holes may be considered up to this precision, with only moderate 
	computational effort.

	In future applications of the scheme we shall calculate a variety of initial 
	data sets. The high accuracy of these data will enable us to investigate 
	physical and mathematical properties thoroughly. Moreover, we intend to 
	study several formulations of the constraint equations in order to select
	physically interesting data which correspond to a binary system 
	with a realistic content of ingoing and outgoing radiation. For this it will 
	prove fruitful to handle explicitly extreme configurations that correspond 
	to limiting cases (such as the test mass limit).
	Finally we plan to use the data in a dynamical evolution of the system which
	ultimately will help clear up the physical significance of the 
	data being considered.  

\begin{acknowledgments}

	The author wishes to thank B.~Br\"{u}gmann for many valuable discussions.
	Moreover he is grateful to D.~Petroff for carefully reading the manuscript.

\end{acknowledgments}



\end{document}